\documentclass[12pt]{article}
\usepackage{amssymb,epsf}
\def\lsim{\mathrel{\rlap {\raise.5ex\hbox{$ < $}}
{\lower.5ex\hbox{$\sim$}}}}
\def\gsim{\mathrel{\rlap {\raise.5ex\hbox{$ > $}}
{\lower.5ex\hbox{$\sim$}}}} 
\def\sqr#1#2{{\vcenter{\vbox{\hrule height.#2pt
        \hbox{\vrule width.#2pt height#1pt \kern#1pt
           \vrule width.#2pt}
        \hrule height.#2pt}}}}

\def\lsim{{\displaystyle
{{\raise-8pt\hbox{$ <$}}
\atop{\raise5pt\hbox{$\sim$}}}}}
\def\gsim{{\displaystyle
{{\raise-8pt\hbox{$ >$}}
\atop{\raise5pt\hbox{$\sim$}}}}}
\def\slsim{{\displaystyle
{{\raise-8pt\hbox{$\scriptstyle <$}}
\atop{\raise5pt\hbox{$\scriptstyle \sim$}}}}}
\def\sgsim{{\displaystyle
{{\raise-8pt\hbox{$\scriptstyle  >$}}
\atop{\raise5pt\hbox{$\scriptstyle \sim$}}}}}
\newskip\humongous \humongous=0pt plus 1000pt minus 1000pt

\newcommand{\sumpf}[0]{\sum_{(H^{\rm f},G^{\rm f})}\! \! \! \!
{\raise
4pt
\hbox{$'$}}\,}

\newcommand{\sump}[0]{\sum_{(H,G)}\! \! {\raise 4pt \hbox{$'$}}\,}

\def\bs{\begin{subequations}}
\def\es{\end{subequations}}
\catcode`\@=11
\newcount\hour
\newcount\minute
\newtoks\amorpm
\hour=\time\divide\hour by60
\minute=\time{\multiply\hour by60 \global\advance\minute by-\hour}
\edef\standardtime{{\ifnum\hour<12 \global\amorpm={am}%
        \else\global\amorpm={pm}\advance\hour by-12 \fi
        \ifnum\hour=0 \hour=12 \fi
        \number\hour:\ifnum\minute<10 0\fi\number\minute\the\amorpm}}
\edef\militarytime{\number\hour:\ifnum\minute<10 0\fi\number\minute}
\def\draftlabel#1{{\@bsphack\if@filesw {\let\thepage\relax
   \xdef\@gtempa{\write\@auxout{\string
      \newlabel{#1}{{\@currentlabel}{\thepage}}}}}\@gtempa
   \if@nobreak \ifvmode\nobreak\fi\fi\fi\@esphack}
        \gdef\@eqnlabel{#1}}
\def\@eqnlabel{}
\def\@vacuum{}
\def\draftmarginnote#1{\marginpar{\raggedright\scriptsize\tt#1}}
\def\draft{\oddsidemargin -.2truein
        \def\@oddfoot{\sl preliminary draft \hfil
        \rm\thepage\hfil\sl\today\quad\militarytime}
        \let\@evenfoot\@oddfoot \overfullrule 3pt
        \let\label=\draftlabel
        \let\marginnote=\draftmarginnote
   \def\@eqnnum{(\theequation)\rlap{\kern\marginparsep\tt\@eqnlabel}%
\global\let\@eqnlabel\@vacuum}  }
\def\subequations{\refstepcounter{equation}%
  \edef\@savedequation{\the\c@equation}%
  \@stequation=\expandafter{\theequation}
  \edef\@savedtheequation{\the\@stequation}
  \edef\oldtheequation{\theequation}%
  \setcounter{equation}{0}%
  \def\theequation{\oldtheequation\alph{equation}}}

\def\endsubequations{\setcounter{equation}{\@savedequation}%
  \@stequation=\expandafter{\@savedtheequation}%
  \edef\theequation{\the\@stequation}\global\@ignoretrue
  \vspace*{-12pt} \\}
\def\bs{\begin{subequations}}
\def\es{\end{subequations}}
\relax
\def\Tr{\,{\rm Tr}\, }

\def\Im{\,{\rm Im}\, }

\def\thefootnote{\fnsymbol{footnote}}
\def\be{\begin{equation}}
\def\ee{\end{equation}}
\def\ba{\begin{eqnarray}}
\def\ea{\end{eqnarray}}

\def\ee{\end{equation}}
\def\bea{\begin{eqnarray}}
\def\eea{\end{eqnarray}}

\newcommand{\uarrw}[0]{\mathrel{
{\raise.5ex\vbox{\hrule width 1cm}\hskip-6pt\rightarrow}}}
\def\thebibliography#1{%
\vskip 0.5cm \centerline{\bf References}
\list{%
[\arabic{enumi}]}{\settowidth\labelwidth{[#1]}
\leftmargin\labelwidth
\advance\leftmargin\labelsep
\usecounter{enumi}}
\def\newblock{\hskip .11em plus .33em minus .07em}
\sloppy\clubpenalty4000\widowpenalty4000
\sfcode`\.=1000\relax}

\renewcommand{\theequation}{\arabic{section}.\arabic{equation}}

\renewcommand{\section}{\setcounter{equation}{0}\@startsection%
{section}{1}{0mm}{-\baselineskip}{0.5\baselineskip}%
{\normalfont\normalsize\bfseries}}

\renewcommand{\subsection}{\@startsection%
{subsection}{2}{0mm}{-\baselineskip}{0.5\baselineskip}%
{\normalfont\normalsize\slshape}}

\renewcommand{\subsubsection}{\@startsection%
{subsubsection}{2}{0mm}{-\baselineskip}{0.5\baselineskip}%
{\normalfont\normalsize\slshape}}

\begin{document}
%
%
\renewcommand{\theequation}{\arabic{section}.\arabic{equation}}
\begin{titlepage}
\begin{flushright}
HU-EP 01/36,\\
hep-th/0110025 
\end{flushright}
\begin{centering}
\vspace{1.0in}

{\bf The ${\bf d=6}$, ${\bf {\cal N}=1}$ heterotic string does 
not ``live'' in six dimensions}$^\dagger$\\
\vspace{1.7 cm}
{\bf{Andrea Gregori}$^1$} \\
\medskip
\vspace{.4in}
{\it  Humboldt-Universit\"at, Institut f\"ur Physik}\\
{\it D-10115 Berlin, Germany}\\

\vspace{2.5cm}
{\bf Abstract}\\
\vspace{.1in}
We discuss why the ${\cal N}_6=1$ heterotic string
has to be viewed as something similar to a   ``non-compact orbifold''. 
Only the perturbative spectrum is forced to satisfy the constraints 
imposed by the vanishing of six-dimensional anomalies. 
These do not apply to the states of the 
non-perturbative spectrum, such as those appearing when small instantons 
shrink to zero size.
\vspace{.1in}
\end{centering}
\vspace{5cm}

\hrule width 3cm
\noindent
$^\dagger$Research supported by the ``Marie Curie'' fellowship
HPMF-CT-1999-00396. Extended version of a work presented at "{\sl Modern Trends in String
Theory}", Lisbon, 13--17 July 2001.
$^1$e-mail: agregori@physik.hu-berlin.de \\

\end{titlepage}
\newpage
\setcounter{footnote}{0}
\renewcommand{\thefootnote}{\arabic{footnote}}

\
\\

The massless spectrum of the heterotic string in ten dimensions 
is determined by the constraint imposed by the vanishing of the Green--Schwarz
anomaly, that requires the presence, together with the ${\cal N}_{10}=1$ 
supergravity multiplet, of 496 gauge bosons; the only two
solutions for the gauge group $G$ are $G= E_8 \times E_8$ and $G= SO(32)$.
The (massless) spectrum of the string in lower dimensions is obtained   
by ``dimensional reduction''. Namely, the degrees of freedom remain the same,
but they are differently interpreted, in terms of lower dimensional fields,
arranged into multiplets of the appropriate ${\cal N}_d$ supersymmetry
in $d$ dimensions.

In less than ten dimensions, the Green-Schwarz anomaly constraint doesn't apply
anymore, and the gauge and matter spectrum can be varied, by introducing 
Wilson lines. In six dimensions, it is also possible to reduce supersymmetry
from the ${\cal N}_6=2$, as derived by toroidal compactification of the 
${\cal N}_{10}=1$, to ${\cal N}_6=1$, by compactifying the four coordinates
on a curved space. As is known, this space is unique, the K3 surface.
Although not uniquely determined, the massless spectrum is nevertheless
highly constrained: it must satisfy a constraint derived by requiring the
cancellation of the ${\cal N}_6=1$, $\Tr R^4$ anomaly.
This constraint reads \cite{schwarz}:
\be
N_H \, - \, N_V \, + \, 29 N_T \; = \; 273 \, .
\label{d6an}
\ee
The ${\cal N}_6=2$ heterotic string has been conjectured to be dual
to the type IIA string compactified on the K3 surface \cite{w}.
When both the theories are further compactified on a two-torus,
$T^2$, this translates into a duality between ${\cal N}_4=4$ theories,
that maps a modulus of the gravity multiplet into a modulus
of the vector manifold \cite{ht}.
The ${\cal N}_6=1$ theory on the other hand does not possess a type IIA
dual. However, when this theory is toroidally compactified on a two-torus,
it is expected to be dual to the type IIA string compactified on
a K3 fibration. The heterotic dilaton--axion field maps then to the 
volume form of the fibration. It may seem a bit strange that, while
the ${\cal N}_4=4$ duality exists also in six dimensions,
this duality of the ${\cal N}_4=2$ theory appears only in four dimensions.
>From a technical point of view, this is related to the fact that it is
not possible to construct an ${\cal N}_6=1$ type II theory.
But it also means that, if string-string duality has to be taken seriously,
the operation of compactification of the ${\cal N}_6=1$ heterotic theory
on $T^2$ is not so an innocuous one: something very special must happen
at the level of the  ``string theory'', 
namely the theory conjectured to be  the basic one, 
underlying all the specific manifestations, whether 
heterotic or type II, or type I string constructions.
>From that point of view, the two-torus must not be a ``flat'' space.

\vspace{.8cm}

In order to understand what is going on, we start by considering in detail
the heterotic/type IIA duality map in the ${\cal N}_4=2$ theory.
The correspondence of the volume form of the base of the type IIA, 
K3 fibration, with the heterotic dilaton can be observed by looking at the
effective theory, built on the massless states of both the constructions. 
The duality map is therefore 
``perturbative/non-perturbative'', and requires for consistency that  
also the modulus parameterizing the coupling of the type IIA string,
belonging to the hypermultiplets, has a heterotic counterpart.
On the heterotic side, the hypermultiplets correspond to moduli in the
K3: the latter must therefore be elliptically fibered, in order to
``contain'' a torus corresponding to the type II coupling, whose
dual must be a perturbative modulus of the heterotic string.
For what matters instead the K3 fibration on the type IIA side,
this also must satisfy certain requirements in order to constitute
a compactification space compatible, in the limit of large volume of the base,
with the heterotic weak coupling limit.
Here however we must clarify certain commonly accepted points that
we find misleading.  
It has often been said that the type IIA space must also be elliptically 
fibered. This property is advocated in order to respect a 
``fiberwise'' identification of the type IIA and heterotic theories
\footnote{See for instance Refs. \cite{aspin,apl}.}.
Namely, one should read the image of the heterotic two-torus
in a torus, appearing, on the type IIA side, as the base of a fibration.
As we will see, the heterotic two-torus cannot be mapped into
a ``torus'' on the type IIA side.
It has also been claimed that the fiber itself should be elliptically
fibered: the fiber is in fact a K3, and it should possess geometric
properties corresponding to those of the K3 on the heterotic side.
This also is not correct: the type IIA compactification space cannot 
``contain'' a torus corresponding to the torus contained in the heterotic 
elliptic K3; this is in fact dual to the type II coupling, 
and the type II string cannot contain as geometric, perturbative modulus, 
its own coupling.
In order to see what should instead be the properties of the type IIA
compact space, we consider a well known example of dual pair,
that of Ref. \cite{fhsv}. In this case, it has been possible to investigate 
both the string theories through the explicit construction of their 
partition function \cite{gkp}, something that allowed to 
perform direct computations of terms of the effective action.
The image of the heterotic two-torus can therefore be followed and checked
explicitly.  

On the type IIA side, the model is obtained by compactification on a
CY$^{11,11}$ manifold. For details about the type IIA and heterotic 
construction, we refer the reader to Refs. \cite{fhsv,gkp}.
Owing to the fact that CY$^{11,11}$ is self-mirror,
both the vector and hyper-multiplets moduli spaces are expected
to be exact, and to not receive neither perturbative nor non-perturbative
corrections. In particular, ${\cal N}_4=2$ supersymmetry is expected to
remain unbroken even non-perturbatively, and we can easily follow the
map of the moduli of the heterotic two-torus and of the dilaton
by comparing, as in Ref. \cite{gkp}, the corrections to the effective
coupling of a term of the effective action, the $R^2$ term.
The reason of this choice is that this quantity receives contributions
only from BPS saturated multiplets, and, owing to unbroken supersymmetry,
the moduli of the heterotic K3 don't enter in the game \cite{hm,hmFHSV}. 
On the heterotic side we can therefore directly compute this quantity at the 
$T^2 \times T^4 \big/ Z_2$ orbifold limit. At this point,
the $Z_2$ acts also as a shift in the two-torus.
Moreover, this orbifold limit turns out to be appropriate because
we are only interested in the map of
the moduli of the torus and of the dilaton, and the $Z_2$ orbifold point
all the Wilson lines are frozen to discrete values. 
As discussed in Ref. \cite{gkp}, this point in the
heterotic moduli space corresponds to the $Z_2 \times Z_2$ 
orbifold point on the type IIA side. The heterotic moduli $T$ and $U$,
associated respectively to the K\"{a}hler class and the complex structure of 
the two-torus, are mapped into the moduli $T^{(i)}$, $i=1,2$,
associated to the K\"{a}hler classes of two of the three 
tori left fixed by the elements of the $Z_2 \times Z_2$ orbifold group.
The K\"{a}hler class of the third torus, $T^{(3)}$, is dual
to the heterotic dilaton: this plane can in fact be seen as
the base of a K3 fibration, whose fiber corresponds to the other two planes.
The model is not symmetric in the three
planes: the first two are acted on by a translation, dual to the
translation acting on the heterotic two-torus (notice that, owing to the
$Z_2 \times Z_2$ action, this space cannot be put in the trivial form
of base$\times$fiber). We learn therefore that, in this particular case: 
\newline
\romannumeral1) the type IIA space respects the $T \leftrightarrow U$
symmetry of the heterotic two-torus, appearing here as a 
symmetry between two fixed planes, $T^{(1)} \leftrightarrow T^{(2)}$,
consequence of the symmetry of the action of two $Z_2$ orbifold elements;
\newline
\romannumeral2) the image of the three heterotic fields $S$, $T$, $U$,
``covers'' the entire type IIA compact space, namely: 
$\langle S \rangle \times \langle T \rangle \times \langle U \rangle
\to \prod_{i=1}^3 \langle T^{(i)} \rangle \approx {\rm Vol}({\rm CY})$.  
This is quite reasonable; the three moduli $S$, $T$, $U$ play in fact
a special role: they must map into the three two-cycles always present in the 
type IIA Calabi--Yau space (see Ref. \cite{al}).

We consider now the decompactification of the heterotic torus.
Here we encounter a first mismatch, in the number of coordinates that
are decompactified on the two sides. Under the heterotic/type IIA map,
the heterotic torus is ``de-constructed'' and ``reconstructed'' back
as a four-submanifold. This ``doubling'' of coordinates is consistent,
because on the heterotic side both the K\"{a}hler class and the
complex structure of the torus belong to the moduli in the vector multiplets,
while on the type II side complex structure and K\"{a}hler class deformations
belong to different moduli manifolds. Therefore, part of the moduli 
of the type IIA four-submanifold will correspond to moduli in the
heterotic K3. Moreover, although the orbifold action is free on both sides,
on the heterotic side the action is trivialized
by decompactifying one coordinate of the torus, while on the type IIA side
this trivialization is achieved only after decompactification of
two coordinates. This also is consistent, because on the heterotic side
the translation in the two torus acts on both the moduli $T$ and $U$,
and the same must be true for its dual image.
This however implies that, while on the heterotic side the orbifold action
is trivialized in five dimensions, on the type IIA side
it is in six dimensions. If we now decompactify the 
heterotic torus, we see that the dual type II picture is ill defined:
if we reach the six dimensional limit by
passing through five dimensions, we take the $T \to \infty$
limit while decompactifying also the field $U$. If instead we
decompactified both the coordinates at the same time, we go to six
dimensions keeping $U$ fixed, or finite.
The decompactification of the heterotic torus may therefore
correspond on the type II side to a decompactification from one up 
to four coordinates: all these situations correspond, on the heterotic side,
to the same limit \footnote{Moreover, we saw that certain moduli in the 
four-submanifold of the type IIA side, that corresponds to the heterotic 
torus, are dual to moduli in the heterotic K3.
Through heterotic/type II/heterotic duality, we see therefore that
the decompactification of the heterotic torus may, by consistency, 
involve also a decompactification of the heterotic K3.}!

>From the point of view of the effective theory,
the ``decompactification'' is quite delicate. In the case of
``freely acting orbifolds'', as is the present case, there are
two decompactification limits: a true, genuine decompactification,
under which the orbifold projection is trivialized,
and a T-dual limit, in which the orbifold operation is not trivialized 
\cite{solving}.
In this second limit we obtain a ``non-compact orbifold'',
namely a space in which certain compact coordinates are ``expanded'',
however, from a global point of view, the space is not a plane but
still an orbifold, with fixed points at which the 
``curvature is concentrated'': we don't have therefore a genuine
six dimensional theory, and in fact the massless degrees of freedom,
once interpreted in terms of states of a six dimensional theory,
do not satisfy the anomaly constraint (\ref{d6an}). The difference between
the two situations is explicit from the very construction of the 
heterotic orbifold, and there is no ambiguity in understanding whether
we are performing a genuine decompactification, as in the first case,
or we are going to the ``non-compact orbifold limit''. 
The translation in the torus, associated to the orbifold projection, 
breaks in fact T-duality explicitly and makes the two limits quite different.
>From the point of view of the underlying theory, in the genuine 
decompactification limit it is not really essential the precise 
identification of the space we end up
on the type IIA side. We know in fact that, under this process, we obtain
an approximate restoration of the initial amount of supersymmetry;
in the case at hand, we get an ${\cal N}_4=4$ theory, in an
appropriate decompactification limit. Since the massless spectrum
of this theory was just a projection of the perturbative
spectrum of the heterotic string, as derived from compactification
of the ten dimensional theory, for what matters the identification of 
the underlying theory we deal with the process of 
compactification/decompactification as in a supergravity theory, 
with projections ``\`{a} la Scherk--Schwarz''.

\vspace{.8cm}

More subtle is the case of a true 
``stringy'' situation. Let's consider a heterotic theory obtained
by compactification on $T^2 \times K3$, with a certain
choice of gauge bundle, but without any explicit action on $T^2$.
Namely, a four dimensional theory obtained just
by simple dimensional reduction on $T^2$ of a six dimensional heterotic
vacuum. In general, this theory possesses a dual realization
as a type IIA string compactified on an appropriate K3 fibration ${\cal M}$.
As before, also in this case the image of the heterotic two-torus must 
``cover'' the fiber, in order to represent, together with the
dilaton field, the three ``minimal'' cycles of the Calabi--Yau ${\cal M}$, 
corresponding to the three ``minimal'' vector fields of this theory.
As before, the decompactification of the heterotic torus doesn't lead
to a smooth space. However, in this case we
cannot use arguments similar to those of freely acting orbifolds,
for which we know that the orbifold projection is trivialized
in the decompactification limit. Here, it is not anymore immaterial
what is the space we end up on the type IIA side, because the final theory
has the same amount of supersymmetry as the four dimensional orbifold.
We are therefore in a situation quite similar to the ``non compact orbifold''
limit of the previous example.
As long as we only consider the perturbative spectrum of the
heterotic string, the compactification on a torus can be handled as in field 
theory: the perturbative spectrum is in fact just the 
``dimensional reduction'' of the one obtained in six dimensions.
Things are however quite different when, at special points
in the hypermultiplets moduli, new massless states
appear, associated to instantons that shrink to zero size \cite{wsi}.
These states are entirely non-perturbative 
from the heterotic point of view. Their existence can however be 
detected on the type IIA dual, where 
they are associated to singular cycles appearing at points in which 
the fiber degenerates \cite{al}.
These states do not disappear when we decompactify the image of the heterotic 
torus, and the ``compactification'' space we end up is not quite a four 
dimensional manifold,
but a singular six dimensional space with some ``directions''
expanded. As it happens for ``non compact orbifolds'', there is no
reason to expect that, once re-interpreted in terms of six dimensions,
the massless states of this theory satisfy the anomaly
constraint (\ref{d6an}), and in fact in general they don't.
In order to see this, we must consider more in detail what is
the structure of the type IIA Calabi--Yau space ${\cal M}$, dual to the 
heterotic string on $T^2 \times K3$. 
This space must possess a symmetry reflecting the 
$T \leftrightarrow U$ symmetry of the heterotic torus. 
Since, on the heterotic side, this symmetry is valid perturbatively, 
and can only be spoiled by instanton corrections of the type 
$ \sim {\rm e}^{-S}$,
it must result on the type IIA side in a \emph{geometric} property of 
the fiber: it can be spoiled only at the level of the quantum theory, by 
exponential terms of the type ${\rm e}^{-n t}$, where $t$ is the modulus
parameterizing the base of the fibration. These terms 
are not contained in the ``geometric'' informations of the Calabi--Yau:
they are proper of the quantum theory, and can be computed using
string--string dualities, such as mirror symmetry \cite{candelas1,candelas2}.
The $T \leftrightarrow U$ symmetry is therefore a symmetry of the 
generic fiber. For what we saw, the
image of the heterotic torus is not a torus of the type IIA side,
but rather something in the fiber of dimension four;
this implies that the  $T \leftrightarrow U$ symmetry is a ``mirror''
symmetry that relates a part of the fiber to another part.
When, by moving in the moduli space, the fiber encounters a singularity,
such that new cycles $E$ appear, intersecting the cycle ``$T$'',
there exists therefore also a ``mirror'' singularity with cycles $E^{\prime}$ 
intersecting ``$U$''. By this we mean that for any new
triple intersection term of the type $E \cap E \cap T$, there is also
a triple intersection $E^{\prime} \cap E^{\prime} \cap U$.
From the heterotic point of view, intersections of this kind
tell us that new vector multiplets appeared, with an effective coupling 
parameterized respectively by the volume of the two-torus, or
its complex structure. Under decompactification to six 
dimensions, all the effective couplings are rescaled by dividing them
by the volume of the two torus, $V_{(2)} \propto T_2$.
This means that terms of the first kind decompactify to six
dimensions to terms with coupling $1$, while terms of the second
kind disappear from the effective action. Being however the two
situations related by T-duality in one coordinate, 
it is easy to realize that the second type of
vectors decompactify to tensors in six dimensions.
We have therefore the following situation: once decompactified to six
dimensions, the perturbative spectrum decompactifies as usual to
a spectrum that satisfies  the constraint (\ref{d6an}). 
The extra states give rise to an equal number of extra vector and tensor 
multiplets. There is no choice of extra hypermultiplets for which
the additional states can satisfy the constraint (\ref{d6an}):
their contribution appears in fact with the same sign as the tensor
multiplets.

Terms with six dimensional coupling $1$ have been investigated 
in several works \cite{mvI,mvII,am}. These analysis of non-perturbative states
of the heterotic string are however incomplete: they neglect the fact
that the ${\cal N}_6=1$ theory is not a ``truly six-dimensional'' theory.
In taking flat limits ($ K3 \to R^4$) on which to study small instantons 
configurations, one always implicitly chooses a preferred direction of 
decompactification, treating the image of the heterotic two-torus on the 
type IIA side as a sub-manifold
of dimension two instead of four. Moreover, any constraint imposed
by the vanishing of six dimensional anomalies cannot be applied,
being completely meaningless because we are not really in six dimensions.
In particular, it is not necessary to introduce couplings of the
extra tensors with Chern--Simons terms, whose existence would be
required by six dimensional anomaly cancellation. 
Analogously, incomplete are also the informations about 
the heterotic non-perturbative spectrum obtained by looking at
the perturbative spectrum of the type I string with D9 and D5 branes.

In our discussion we made use of string--string duality in order to make
clear that the ${\cal N}_6=1$ heterotic string
doesn't really live in six dimensions.
However, we don't really need heterotic/type II duality in order
to see that the non-perturbative spectrum violates the constraint (\ref{d6an}).
This can be seen already from the heterotic string itself.
Let's in fact suppose that, at certain points in the K3 moduli space
new massless states appear, with coupling 1 in six dimensions.
Under compactification to four dimensions, their effective coupling
becomes then $1 \big/ g^2_{(4)} \propto V_{(\rm torus)} \equiv \Im T$.
The absence of ``mirror'' states with coupling $\Im U$ explicitly
contradicts the $T \leftrightarrow U$ symmetry, direct consequence of
T-duality and toroidal compactification. The six dimensional theory
must therefore contain also these states, that can only be tensors;
(\ref{d6an}) is then automatically violated. We could on the other hand
suppose that T-duality is preserved because, 
under compactification, in some misterious way the 
coupling ``1'' of six dimensions acquires, besides the volume factor 
$\Im T$, also a dependence on $U$: $1 \big/ g^2_{(4)} \approx \Im T + \Im U$
(to be eventually promoted to an $SL(2,Z)$-invariant function of these 
fields). Such an expression is compatible with the coupling of six
dimensions, because the $U$-dependent part vanishes anyway under
decompactification and rescaling with the volume of the two-torus.
However, with such a coupling, from the four dimensional point of view
we could not discriminate whether the new vector fields
originate from six dimensional vectors or tensors: we would 
have an effective doubling of kinetic and interaction terms, and
therefore get anyway into a contradiction. In summary, 

\emph{perturbative T-duality and the appearance of non-perturbative 
massless states necessarily implies that the six dimensional anomaly constraint
(\ref{d6an}) is violated}.

We stress that this is peculiar of string theory, and
does not happen in field theory/supergravity: T-duality and
non-perturbative states of the ``small instantons'' type are pure
stringy phenomena.

As a last comment, we remark that, always by string-string duality, we learn
that also the ten dimensional ${\cal N}_{10}=1$ heterotic string
is in some sense a ``non-compact orbifold''. 
This can be seen also by considering it as a Ho\v{r}ava--Witten orbifold:
when a further circle of the M-theory is compactified to a small radius, we
fall into a perturbative type IIA string situation, for which
it is not possible to project with a $Z_2$ twist only one coordinate.
However, in this case there are no non-perturbative
extensions of the heterotic massless spectrum; 
therefore, one can safely neglect 
such details and deal with the string massless spectra as in field theory.

\vspace{1.5cm}


\providecommand{\href}[2]{#2}\begingroup\raggedright\endgroup

\bibliography{robert}
\bibliographystyle{ieert}

\end{document}